# Controlling the polarisation correlation of photon pairs from a charge-tuneable quantum dot

R. J. Young,[a)] S. J. Dewhurst, R. M. Stevenson, A. J. Shields
*Toshiba Research Europe Limited, 260 Cambridge Science Park, Cambridge CB4 0WE, UK*

P. Atkinson, K. Cooper, D. A. Ritchie
*Cavendish Laboratory, University of Cambridge, Cambridge CB3 0HE, UK*

a)Electronic mail: robert.young@crl.toshiba.co.uk

Correlation between the rectilinear polarisations of the photons emitted from the biexciton decay in a single quantum dot is investigated in a device which allows the charge-state of the dot to be controlled. Optimising emission from the neutral exciton states maximises the operating efficiency of the biexciton decay. This is important for single dot applications such as a triggered source of entangled photons. As the bias on the device is reduced correlation between the two photons is found to fall dramatically as emission from the negatively charged exciton becomes significant. Lifetime measurements demonstrate that electronic spin-scattering is the likely cause.

Single quantum dots can confine excitons in all three spatial dimensions[1] providing a useful system for generating non-classical light. Semiconductor fabrication allows quantum dots to be integrated into device structures such as diodes[2] and optical cavities[3] promising the realisation of compact and robust light sources which could be useful for quantum information processing. Single photon emission has been extensively studied from various exciton states confined by quantum dots[4,5]. Recently polarisation-entangled photon pair emission from the biexciton decay in a single dot was shown[6,7], providing a semiconductor source of triggered entangled photons.

Photoluminescence spectra from single quantum dots typically contain multiple emission lines corresponding to recombination of excitons with more than one charge state[8]. This is highlighted in figure 1, in which emission from both neutral exciton complexes (X and XX) and the positively charged exciton ($X^+$) are clearly visible. The presence of emission from exciton complexes of differing charge composition is a result of random migration of electrons and holes into the quantum dot. This switching of the exciton charge state in single quantum dots is undesirable for many quantum dot applications. An example of this is the biexciton decay which is used as a triggered source of photon pairs; the emission cycles in which charged excitons form prevent emission from the biexciton decay and directly limit the operating efficiency of a single dot based device.

To enable the exciton charge state of individual dots to be controlled they were placed in the intrinsic region of a p-i-n diode structure. Small self-assembled quantum dots were grown by molecular beam epitaxy; a layer of InAs was grown on GaAs close to the critical thickness for dot formation. A planar optical cavity[9] at ~900nm was integrated into the device to increase the proportion of the light emitted by the quantum dot that was collected. This was achieved by growing 12 repeats of alternating quarter-wavelength thick layers of GaAs/AlAs distributed Bragg reflectors below the $\lambda$-cavity containing the dot layer, and two repeats above the dot layer. The top two layers of the bottom mirror were doped n-type with silicon and the top mirror was p-doped with carbon.

The dot layer was situated on top of a 5nm layer of GaAs above an $Al_{0.98}Ga_{0.02}As/Al_{0.5}Ga_{0.5}As$ superlattice formed with 7 repeats of 2nm thick layers. The superlattice barrier prohibits electrons from tunnelling out of the quantum dots. To isolate single quantum dots and facilitate easy relocation a metal shadow mask was placed on top of the sample containing ~2µm diameter circular apertures.

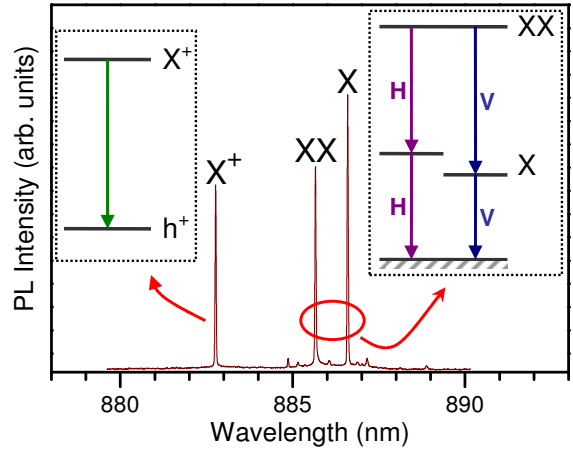

FIG. 1. Photoluminescence measured from a single quantum dot. The three sharp peaks correspond to emission from the neutral exciton (X) and biexciton (XX) and the positively charged exciton ($X^+$) states, as labelled. The simplified energy level diagrams illustrate the initial and final states involved in the main emission lines ($h^+$ represents a single hole). The photons from the biexciton decay are typically horizontally (H) and vertically (V) polarised.

The device was cooled to <10K in a continuous flow Helium-4 cryostat. Photoluminescence was collected by a microscope objective and recorded with a grating spectrometer. A pulsed laser diode operating at 80MHz with pulse duration of ~100ps was used to excite the sample above the GaAs bandgap.

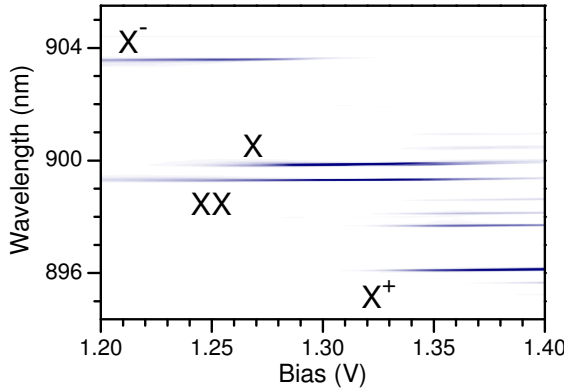

FIG. 2. Photoluminescence from a single quantum dot embedded in a p-i-n structure as a function of the bias applied between the n- and p-contacts. Blue areas indicate high count rates and white areas indicate low count rates.

Photoluminescence from a single quantum dot was measured as a function of the bias applied between the p- and n-type regions of the device in figure 2. Below a bias of ~1V no photoluminescence was measured from the quantum dots indicating that the rate at which the heavy-holes tunnel from the dots was much greater than the exciton's radiative lifetime. At 1-1.2V photoluminescence was predominantly measured from the negatively charged exciton ($X^-$). As the bias was increased further emission from the neutral exciton (X) and biexciton (XX) states became dominant followed by emission from the positively charged exciton ($X^+$). At larger biases a number of weak emission lines appeared in the photoluminescence thought to be from higher positively charged states. This demonstrates that the device can be used to carefully control the charge of the exciton complexes which are dominant in the photoluminescence spectra of a single quantum dot[10].

The intensities of the emission lines from each of the four main exciton states in figure 2 were integrated and are shown as a function of bias in figure 3 (a). This shows that in the bias range 1.30-1.32V the emission is almost entirely from the neutral biexciton decay.

The quantum dot studied here has a significant fine structure splitting[11] (>20μeV) between its bright exciton states. It is expected therefore that the photons from its biexciton decay will be classically polarisation-correlated in a rectilinear (vertical V and horizontal H) basis[12] as illustrated in the level diagram shown in the inset on the right of figure 1. To investigate the influence of the applied bias on this correlation a second spectrometer was used to allow the exciton and biexciton photons to be spectrally separated. The exciton photons were selected with the first spectrometer and a linearly polarising beam splitter passed the emission into a pair of Avalanche PhotoDiodes (APD's). The second spectrometer was set to filter biexciton photons which were linearly polarised and passed into a single APD. The time delay between measuring a biexciton photon and each of the two rectilinear polarised exciton photons was measured simultaneously. These two measurements were used to calculate the degree of rectilinear polarisation correlation (C) between the biexciton and exciton photons. This is defined as $C = \frac{g_{xx,x} - g_{xx,\bar{x}}}{g_{xx,x} + g_{xx,\bar{x}}}$ where the second order correlation functions $g_{xx,x}$ and $g_{xx,\bar{x}}$ are the simultaneously measured, normalised coincidences of the horizontally polarised XX photon with the co-polarised X and vertically polarised $\bar{X}$ photons respectively. C is therefore expected to be 100% for an ideal unpolarised source emitting polarisation correlated photons in the rectilinear basis, and 0% for an uncorrelated source. C was measured with a number of different biases applied to the device; these results are shown in the figure 3(b).

Studies of nominally uncharged quantum dots have shown that the degree of correlation is limited by background emission due to emission from layers other than the dot, as well as exciton spin scattering[7]. The former mechanism will be least significant when the XX/X photon pair rate from the dot is maximal. This occurs for a bias voltage of ~1.31V in figure 3 (a). Notice, however, that the degree of correlation in figure 3 (b) increases at higher biases from 56±4% at 1.31V to 68±6% at 1.33V. Furthermore the correlation drops sharply at lower voltages to 13±5% at 1.25V. This sharp degradation is likely to be due to exciton spin scattering, induced by excess electrons in the vicinity of the dot, as now discussed.

The exciton and biexciton state lifetimes were measured as a function of bias. Time resolved emission from the two states was measured as before but now the time interval between the laser trigger and photon detection was recorded. Lifetimes were extracted from the decay curves produced. The ratio of the biexciton's lifetime to the exciton's lifetime is plotted as a function of bias in figure 3 alongside the degree of correlation C. At high biases the polarisation correlation between the pair of photons is high and the ratio of lifetimes is close to 2. This result is expected in the limit of slow exciton dephasing[13,14] and is the typical ratio of lifetimes observed in InAs quantum dots. As the applied bias is reduced, and the degree of correlation drops, the lifetime ratio is found to peak at 3.62±0.02. Theoretical studies of the lifetime ratio predict a value of ~4 in the fast spin-flip limit[14]. A long-lived dark exciton state is formed when the electron in an optically active exciton state undergoes a spin-flip. Such events would therefore increase the measured lifetime of the exciton and explain the large drop in correlation measured. At the lowest biases the lifetime ratio is found to decrease again, as the hole tunnelling time out of the dot becomes comparable to the exciton's radiative lifetime (~1ns).

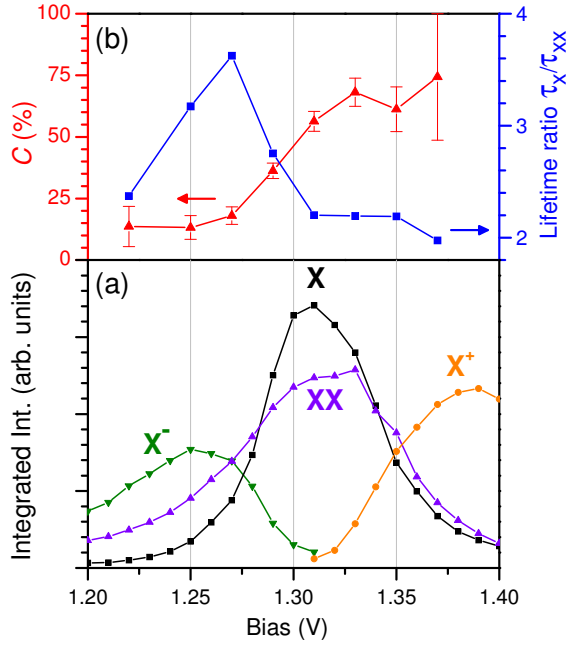

FIG. 3. Plotted as a function of the bias applied to the device: (a) the integrated intensities of the photoluminescence from the negatively charged (X-), neutral (X), bi-(XX) and positively charged exciton (X+) states. (b) The ratio of the exciton and biexciton states' lifetimes and the degree of rectilinear polarisation correlation C between the biexciton and exciton photons. Error bars span 2 standard deviations of the random error. Where not shown the random error is smaller than the symbol size.

In summary we have shown how the polarisation correlation between biexciton and exciton photons varies in a device in which the charge state of the dominant exciton complex can be finely controlled. Choosing a bias to minimise emission from states with mismatched numbers of electrons and holes maximises emission from the neutral exciton as expected. With the charge-state in the regime of the average number of electrons being greater than the average number of holes, spin-scattering severely limits the degree of polarisation correlation measured between the biexciton and exciton photons. It seems likely that this is due to the flipping of the exciton spin by excess electrons in close proximity to the dot. These excess electrons may be confined at the heterojunction formed at the electron tunnel barrier in the layer structure. In contrast, a similar loss of polarisation correlation is not observed at the biases for which an excess of holes around the dot is expected. This may be because hole-exciton spin scattering is weaker than electron-exciton, or because the density of excess holes is lower due to the absence of a tunnel barrier for the holes.

An interesting feature of this device is that the bias required to maximise the pair rate differs slightly from the bias under which the largest proportion of pairs are polarisation-correlated. Choosing a bias between these two values optimises the two-photon emission from the quantum dot. This is highly desirable for devices based on the biexciton decay, such as an entangled photon-pair source.

This work was partially funded by the EU projects QAP and SANDiE, and by the EPSRC through the IRC for Quantum Information Processing.

References:

[1] D. Bimberg, M. Grundmann, and N. N. Ledentsov, *Quantum Dot Heterostructures*. (Wiley, Chichester, 1999).

[2] Z. Yuan, B. E. Kardynal, R. M. Stevenson, A. J. Shields, C. J. Lobo, K. Cooper, N. S. Beattie, D. A. Ritchie, and M. Pepper, Science **295**, 102 (2001).

[3] J. P. Reithmaier, G. Sek, A. Löffler, C. Hofmann, S. Kuhn, S. Reitzenstein, L. V. Keldysh, V. D. Kulakovskii, T. L. Reinecke, and A. Forchel, Nature **432**, 197 (2004); T. Yoshie, A. Scherer, J. Hendrickson, G. Khitrova, H. M. Gibbs, G Rupper, C. Ell, O. B. Shchekin, and D. G. Deppeet, Nature **432**, 200 (2004); E. Peter, P. Senellart, D. Martrou, A. Lemaître, J. Hours, J. M. Gérard, and J. Bloch, Phys. Rev. Lett. **95**, 067401 (2005).

[4] P. Michler, A. Imamoğlu, M. D. Mason, P. J. Carson, G. F. Strouse, and S. K. Buratto, Nature **406**, 968 (2000); V. Zwiller, T. Aichele, W. Seifert, J. Persson, and O. Benson, Appl. Phys. Lett. **82**, 1509 (2003).

[5] R. M. Thompson, R. M. Stevenson, A. J. Shields, I. Farrer, C. J. Lobo, D. A. Ritchie, M. L. Leadbeater, and M. Pepper, Phys. Rev. B **64**, 201302 (2001).

[6] R. M. Stevenson, R. J. Young, P. Atkinson, K. Cooper, D. A. Ritchie, and A. J. Shields, Nature **439**, 179 (2006).

[7] R. J. Young, R. M. Stevenson, P. Atkinson, K. Cooper, D. A. Ritchie, and A. J. Shields, New J. Phys. **8**, 29 (2006).

[8] L. Landin, M. S. Miller, M.-E. Pistol, C. E. Pryor, and L. Samuelson, Science **280**, 262 (1998); A. Hartmann, Ducommun. Y., E. Kapon, U. Hohenester, and E. Molinari, Phys. Rev. Lett. **84**, 5648 (2000).

[9] H. Benisty, H. De Neve, and C. Weisbuch, IEEE J. Quantum Electron **34**, 1612 (1998).

[10] R. J. Warburton, C. Schäflein, D. Haft, F. Bickel, A. Lorke, K. Karrai, J. M. Garcia, W. Schoenfeld, and P. M. Petroff, Nature **405**, 926 (2000); J. J. Finley, P. W. Fry, A. D. Ashmore, A. Lemaître, A. I. Tartakovskii, R. Oulton, D. J. Mowbray, M. S. Skolnick, M. Hopkinson, P. D. Buckle, and P. A. Maksym, Phys. Rev. B **63**, 161305 (2001).

[11] D. Gammon, E. S. Snow, B. V. Shanabrook, D. S. Katzer, and D. Park, Phys. Rev. Lett. **76**, 3005 (1996).

[12] C. Santori, D. Fattal, M. Pelton, G. Solomon, and Y. Yamamoto, Phys. Rev. B **66**, 045308 (2002); R. M. Stevenson, R. M. Thompson, A. J. Shields, I. Farrer, B. E. Kardynal, D. A. Ritchie, and M. Pepper, Phys. Rev. B **66**, 081302 (2002); S. M. Ulrich, S. Strauf, P. Michler, G. Bacher, and A. Forchel, Appl. Phys. Lett. **83**, 1848 (2003).

[13] M. Wimmer, S. V. Nair, and J. Shumway, Phys. Rev. B **73**, 165305 (2006).

[14] G. A. Narvaez, G. Bester, A. Franceschetti, and A. Zunger, Phys. Rev. B **74**, 205422 (2006).